\documentclass{nature}

\usepackage{graphicx} 
\usepackage{lineno}

\makeatletter
\def\makeLineNumberLeft{%
  \linenumberfont\llap{\hb@xt@\linenumberwidth{\LineNumber\hss}\hskip\linenumbersep}
  \hskip\columnwidth
  \rlap{\hskip\linenumbersep\hb@xt@\linenumberwidth{\hss\LineNumber}}\hss}
\leftlinenumbers
\makeatother

\def\gtrsim{\mathrel{\hbox{\rlap{\hbox{\lower5pt\hbox{$\sim$}}}\hbox{$>$}}}}
\newcommand{\msun}{\mbox{M$_\odot$}}
\newcommand{\rsun}{\mbox{R$_\odot$}}
\newcommand{\lsun}{\mbox{L$_\odot$}}
\newcommand{\jj}[2]{\mbox{$J = #1\rightarrow#2$}}
\newcommand{\cooo}{\mbox{C$^{18}$O}}

\newcommand{\arcsec}{\mbox{$^{\prime \prime}$}}
\newcommand{\degree}{\mbox{$^\circ$}}

\title{Formation of Wide Binaries by Turbulent Fragmentation}

\author{Jeong-Eun Lee$^{1}$, Seokho Lee$^1$, Michael Dunham$^{2}$, Ken'ichi Tatematsu$^{3}$, Minho Choi$^4$, Edwin A. Bergin$^5$,
\& Neal J. Evans II$^{4,6}$ }

\begin{document}

\maketitle

\begin{affiliations}
 \item School of Space Research, Kyung Hee University, 1732, Deogyeong-Daero, Giheung-gu, Yongin-shi, Gyunggi-do 17104, Korea
 \item State University of New York at Fredonia, Fredonia, 280 Central Ave., NY 14063, USA
 \item National Astronomical Observatory of Japan, 2-21-1 Osawa, Mitaka, Tokyo 181-8588, Japan
\item Korea Astronomy and Space Science Institute, 776 Daedeokdae-ro, Yuseong-gu, Daejeon, 34055,  Korea
 \item Department of Astronomy, University of Michigan, 500 Church St., Ann
Arbor, MI 48109, USA
  \item Department of Astronomy, The University of Texas at Austin, 
   2515 Speedway, Stop C1402, Austin, TX 78712, USA

 \end{affiliations}


\begin{abstract}
Understanding the formation of  wide binary systems of very low mass stars ($M\le 0.1$ \msun)
is challenging\cite{Bejar2008,Luhman2009,Luhman2012}.
The most obvious route is via widely separated low-mass collapsing fragments produced 
through turbulent fragmentation of a molecular core\cite{Goodwin2004, Fisher2004}. 
However, close binaries/multiples from disk fragmentation can also evolve to wide binaries
over a few initial crossing times of the stellar cluster through tidal evolution\cite{Marks2012}.
Finding an isolated low mass wide binary system in the earliest stage of formation, before tidal evolution could 
 occur,
would prove that  turbulent fragmentation is a viable mechanism for (very) low mass wide binaries. 
Here we report high resolution ALMA observations of a known wide-separation protostellar binary, showing that each component has a circumstellar disk.
The system is too young\cite{Dunham2006} to have evolved from a close binary and the disk axes are misaligned, providing strong support for the turbulent
fragmentation model. Masses of both stars are derived from the Keplerian
rotation of the disks; both are very low mass stars.

\end{abstract}

The two main theoretical formation mechanisms for multiple stars are turbulent fragmentation and disk fragmentation.
These mechanisms result in different distributions of companion separations; wide (separation $>$ 500 AU) binaries naturally form
through turbulent fragmentation\cite{Offner2010}, and close binaries form through disk fragmentation\cite{Kratter2010}.
However, dynamical processes
such as three-body interactions, radial migrations,
and interactions with cluster members\cite{Bate2012,Offner2010,Marks2012}
can turn initially close binaries into wide binaries.
Therefore, in order to demonstrate formation by turbulent fragmentation,
it is crucial to observe protostellar binary or multiple systems,
which are younger than 0.5 Myrs, 
the timescale of the embedded protostellar phase\cite{Dunham2014}, 
and thus, can reveal their initial configurations.

A survey of protostars in the Perseus molecular cloud
found a bi-modal distribution of companion separations
with a peak at 75 AU and another at 3000 AU\cite{Tobin2016}. 
Therefore, evidence exists for both modes of binary formation.
In addition, recent high resolution observations isolate specific targets to support each formation mechanism.
Compelling evidence for disk fragmentation was revealed by a triple
protostellar system forming in a common protostellar disk\cite{Tobin2016b}.
On the other hand, a quadruple condensation system was discovered
and the authors claimed that it is evolving into a stellar system with a wide separation\cite{Pineda2015};
however, this system has so far formed only one protostar, and its ultimate outcome is unclear.

Although the systems with larger separations
are plausible candidates for formation by turbulent fragmentation,
confirmation of this possibility requires further evidence,  {\it the misalignment of rotation axes} of stars, disks, and outflows\cite{Offner2016}.
Turbulent fragmentation generates binaries with misaligned rotation axes
because the angular momentum distribution in a turbulent core
is random\cite{Offner2010}.
By contrast, binaries with aligned rotation axes ($\le 20\degree$)\cite{Offner2016} are predicted
if the secondary member is formed in a large co-rotating massive disk/ring
around the primary\cite{Bonnell1994}
or if they are formed through the fragmentation
driven by the centrifugal force in a flattened cloud core.
 Protostellar rotational axes are difficult to measure, but the bipolar outflow driven by a star-disk system
can be used to test the alignment of angular momentum in the system\cite{Offner2016}. 
However, outflow studies suffer from environmental effects\cite{Lee2017} or interactions between multiple outflows. 
A better indicator is the misalignment of disk rotation axes, but disk rotation axes of wide binaries 
have been reported only in the T-Tauri stage (more evolved than the protostellar stage)
\cite{Jensen2004,Jensen2014,Salyk2014,Williams2014},
which suffer from tidal evolution  effects.

IRAS 04191+1523 is a known protostellar binary system
with a projected separation of 6$''$.1 ($\sim 860$ AU) between members\cite{Duchene2004} in Taurus,
whose distance is 140 pc.
Both companion protostars are in the early embedded evolutionary stage of Class I\cite{Dunham2006,Luhman2010}.
The primary (protostar A) has a luminosity of 0.6 \lsun,
and the secondary (protostar B) has
0.04 \lsun\cite{Duchene2004,Connelley2008,Dang-Duc2016}.
Based on their luminosities, the secondary was considered as a candidate proto-brown dwarf\cite{Bulger2014}.
However, the protostellar luminosity is dominated by accretion luminosity,
which depends on both stellar mass and accretion rate. 
The only accurate way to compute protostellar masses is to observe
Keplerian rotation\cite{Tobin2012}.

\begin{figure*}
\includegraphics[height=10cm]{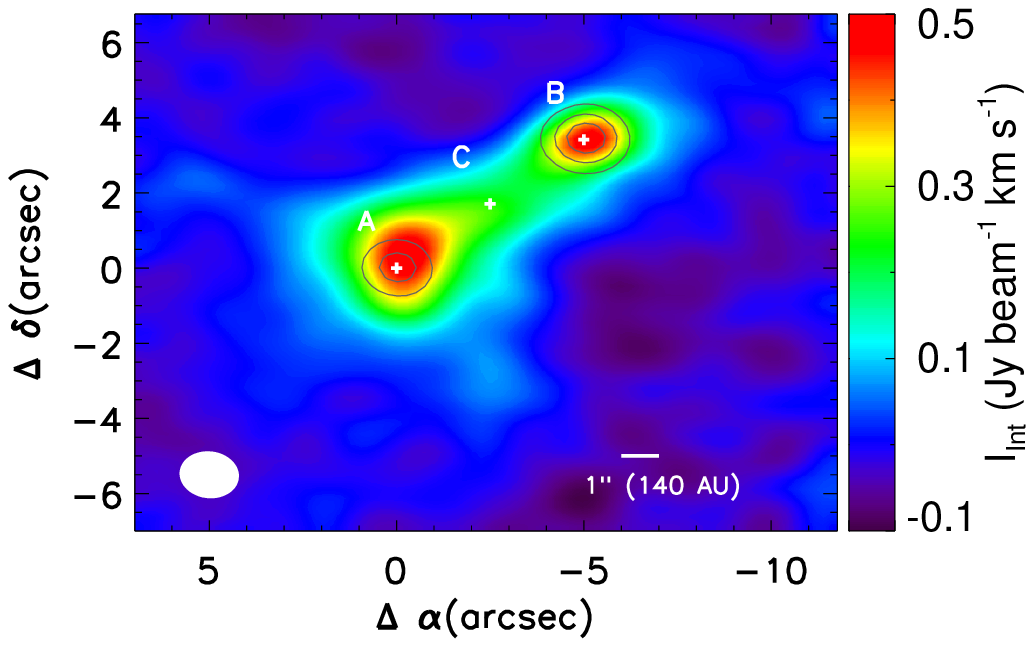}
\caption{Two compact continuum sources detected in the 1.3 mm continuum (black contours) on top of the \cooo\ J=2-1 integrated intensity (colour image). The lowest contour and subsequent contour step are 20 times  the rms noise of 1.0 mJy beam$^{-1}$. A and B indicate the positions of protostars. More extended \cooo\ emission exists around A although the 1.3 mm continuum emission is stronger around B. C represents a common envelope position not associated with compact continuum emission. The peak intensities (and their 1 $\sigma$ error) of the two continuum sources, which are measured with the task $imfit$ in CASA, are 51 ($\pm0.9$) and 76 ($\pm1.0$)  mJy beam$^{-1}$, respectively. }
\label{fig:c18o}
\end{figure*}

IRAS 04191+1523 was observed with ALMA simultaneously in the 1.3 mm continuum and the \cooo\ (J=2--1) emission on 4 January 2015, with $\sim$1.5\arcsec\ resolution.
The 1.3 mm continuum was re-observed with ALMA, obtaining better resolution ($\sim$0.2\arcsec) on 28 September 2016.
The ALMA observations (Figure \ref{fig:c18o}) at the lower resolution reveal
two compact continuum sources, one for each protostar,
connected by a bridge of emission in \cooo. 
The \cooo\ gas bridge and the same central velocities inferred from the \cooo\ spectra over the elongated feature show
that the two protostars belong to a single star-forming cloud core. 
 The Herschel/PACS images\cite{Bulger2014} show a single core, flattened along the this \cooo\ 
emission bridge, with a size of ~0.02 pc, demonstrating sub-fragmentation within an isolated core.

\begin{figure*}
\includegraphics[height=15cm]{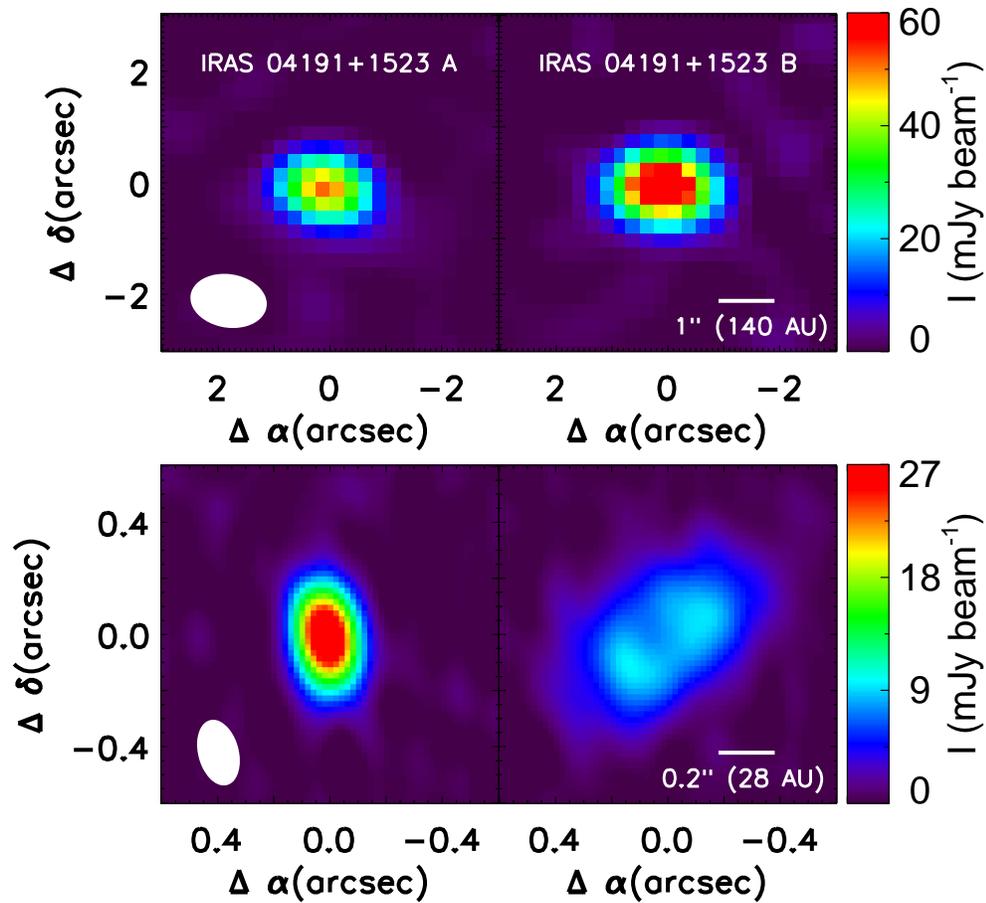}
\caption{Continuum images in two different resolutions. The upper and lower panels show the 1.3 mm continuum images 
in the resolutions of $\sim$1.2\arcsec\ and $\sim$0.2\arcsec, respectively, when the uniform weighting is adopted. 
In the higher resolution images, two continuum sources are resolved as disks (see the text and Supplementary Table 1 and 2).
}
\label{fig:cont}
\end{figure*}

The two continuum sources were resolved only in the higher resolution image (Figure \ref{fig:cont}). 
In the unresolved images (upper panel of Figure 3), 
the total continuum fluxes are 52.6 ($\pm~1.6$) mJy and 87.1 ($\pm~2.0$) mJy for protostars A and B, respectively. 
In the resolved images (lower panel of Figure 3), 
they are 53.5 ($\pm~1.1$) mJy and 71.1 ($\pm~3.2$) mJy for protostars A and B, respectively.
The agreement between continuum fluxes from the two resolution images of protostar A indicates 
that the continuum sources are disks. 
On the other hand, the difference in flux between the two different images of protostar B indicates 
that no more than 20\% (16/87) can be attributed to an envelope. 
We assume that the flux from the disk is 71.1 mJy as measured from the higher resolution image.

We use the higher resolution continuum image to derive the physical parameters of the continuum sources;
the continuum comes from the thermal radiation of dust and provides the masses of circumstellar material.
Each of the dust continuum sources has the size and mass of a typical protostellar accretion disk (see Methods).
The projected position angles of two disks derived from the dust continuum emission are significantly different by $55\pm 5$ degrees
(Supplementary Table 1), indicative of misaligned disk axes.

\begin{figure*}
\includegraphics[height=10cm]{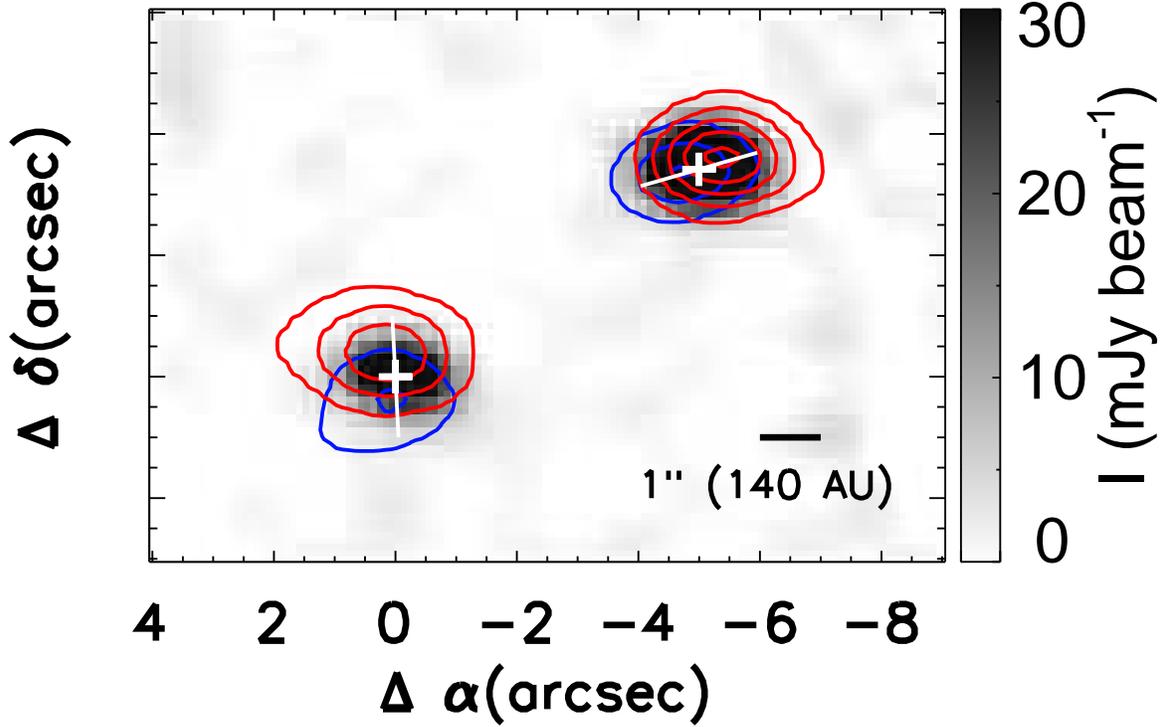}
\vskip -0.5cm
\caption{Protostellar binary disks detected in the high velocity wings of the \cooo\ \jj21\ emission line 
on top of the continuum image (gray scale). 
The red and blue components are integrated in the velocity ranges from 7.65 to 9.15 km~s$^{-1}$ and 
from 4.35 to 5.85 km~s$^{-1}$, respectively. 
The lowest contour and subsequent contour step are 10 times the rms noise of 2.4 mJy beam$^{-1}$ km s$^{-1}$.
The white lines connect the blue- and red-shifted emission peaks, and thus, they are considered to be perpendicular 
to the disk rotation axes. The peak-position-velocity diagrams, presented in Figure 4, are obtained along these lines.
}
\label{fig:disk}
\end{figure*}

\begin{figure*}
\includegraphics[height=7cm]{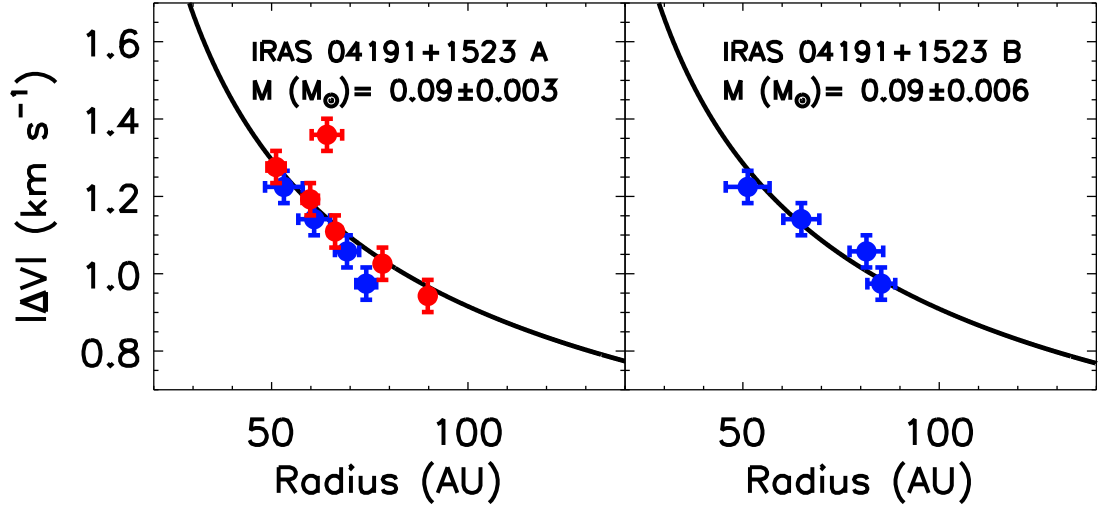}
\vskip -0.5cm
\caption{The velocity profiles for high velocity wings of IRAS 04191+1523 A and B, respectively, 
along the white lines in Figure 3. 
Only blue shifted velocities are plotted for B because the red shifted emission is contaminated 
by an additional component (see Supplementary Figure 1). 
The source velocity is found from the spectrum at the position C in Figure 1 (see Supplementary Figure 1), 
as 6.75 km~s$^{-1}$, in which the velocity profiles for red and blue-shifted 
wings of IRAS 04191+1523 A are consistent without displacement as seen in the figure.
The solid curves represent the best-fit Keplerian velocity profiles with the central mass of 0.09 M$_\odot$ 
for both sources when an edge-on view is assumed. 
The vertical error bars represent the velocity resolution ($\sim$0.08 km s$^{-1}$), and 
the horizontal error bars indicate the 1 $\sigma$ error in the position of emission peak, which is derived by 
Gaussian fitting of the C$^{18}$O emission distribution along the white lines marked in Figure 3 at a given velocity channel.}
\label{fig:kepler}
\end{figure*}

Since the gaseous disks are resolved even at the lower resolution, 
the C$^{18}$O emission provides information
on the kinematics of the disks;
the \cooo\ emission associated with each disk shows
a velocity shift from one end of the disk to the other end,
indicating rotation (Figure \ref{fig:disk}).
In the red-shifted and blue-shifted velocity components together,
the velocity shift is consistent with a Keplerian rotation profile (see Methods).
Figure \ref{fig:kepler} presents the best-fit Keplerian rotation velocity profile in the peak-position-velocity diagram
of each source; the position of the emission peak  is found by Gaussian fitting of the emission
distribution along the white line in Figure \ref{fig:disk} for a given velocity.
The minimum masses of the central objects (the protostars)
can be derived from the best-fit models of these peak-position-velocity diagrams.
Protostar A and B both have $M_*=0.09$ \msun, assuming edge-on disks (see Supplementary Table 3).
If the disk has an intermediate inclination,
the corresponding mass should be scaled by $1/{\rm sin}^2i$,
where $i$ is the inclination angle ($i=90\degree$ for edge-on disk).
The aspect ratio of deconvolved continuum source size suggests
the inclination angles of  55\degree\ and 59\degree\ for the disks of protostars A and B, respectively (see Methods). 
Therefore, the masses after correction for the inclinations are $\sim$0.14 and 0.12 \msun,  which are close to
the criterion of very low mass star ($M_*\le 0.1$ \msun). 

Protostar B cannot be substellar, unlike what would be expected from its very low luminosity\cite{Bulger2014}.
Assuming that most of the luminosity
comes from the mass accretion onto the protostars and protostellar radii are 2\rsun,
the accretion rate of protostar A ($\dot{M}\sim2.8\times 10^{-7}$ \msun yr$^{-1}$)
is greater than that of protostar B ($\dot{M}\sim2.1\times 10^{-8}$ \msun yr$^{-1}$)
by a factor of $\sim13$.
The current protostellar masses are similar to those of very low-mass stars,
but they may grow by further accretion from the disks and envelope.
Assuming a dust temperature of 30 K and a gas-to-dust mass ratio of 100, 
the disk masses are $(3.70 \pm 0.08) \times 10^{-3}$ \msun\
and $(4.91\pm 0.22) \times 10^{-3}$ \msun\ for protostars A and B,
respectively (see Methods). 
These disks cannot substantially increase the central masses. 
The total mass of the protostellar envelope is 0.1 \msun\cite{Young2006}.
Assuming that 30\%--50\% of the envelope mass
eventually winds up in the protostars\cite{Dunham2014},
the future contribution from the envelope may be $\sim$0.04 \msun.
Consequently, the protostars cannot grow very much,
and the final configuration is likely to be a pair of low mass stars
separated by at least 1000 AU.
If the accretion rate of protostar A
stays much larger than that of protostar B even in the future,
the primary mass will be about 0.18 \msun, which is not much bigger than the secondary mass (0.12 \msun).

From Kepler's third law, with the total mass of 0.26 \msun, 
the orbital period of the binary system is $\sim5\times 10^4~(\pm~5.5\times10^3)$ years 
if we consider the projected separation (860 AU) as the actual separation between companions; the dominant error of 
orbital period is the distance error, which is about 10 pc.
If the separation on the plane of sky is the same as the separation along the line of sight, 
the actual separation is $\sqrt2$ times the projected separation, and the orbital period becomes 0.1 Myr.
A significant change of system configuration by tidal interactions
would take a time much longer than the orbital period.
The age of IRAS 04191+1523 is about 0.1 Myr in the assumption of 
a typical protostellar accretion rate of $10^{-6}$ \msun yr$^{-1}$ and the protostellar mass of 0.1 \msun.
Its age should not exceed the timescale (0.5 Myr) of the protostellar phase\cite{Dunham2014}.
 The similarity between the envelope and each protostellar mass indicates that IRAS 04191+1523 is a relatively young 
Class I source\cite{Andre1999}. Therefore, the orbital period of the IRAS 04191+1523 system is comparable
with its age, and this binary system should still be in its initial configuration.

The projected disk rotation axes are perpendicular to the white lines over two gaseous disks in FIgure \ref{fig:disk}, 
which are misaligned by 77\degree\ (see Methods), indicating that they formed via turbulent fragmentation 
since this binary is too young to modify the alignment of rotational axes by tidal interactions between companions\cite{Offner2016}.
The derived orbital period of the isolated low mass protostellar binary system, IRAS 04191+1523, 
also demonstrates that wide binary systems can be formed by the turbulent fragmentation mechanism,
without disk fragmentation/migration.
Disk fragmentation and tidal evolution by cluster members cannot explain the existence of this pair either
because the stellar density in Taurus is too low\cite{Rebull2010} to alter their orbital parameters. 
The timescale for the tidal evolution in Taurus is on the order of  a few crossing times\cite{Marks2012}, 
or about 100 Myr, much longer than the age of this system. 
The crossing time for a low density cluster with the stellar initial half-mass radius of 2.53 pc 
and the initial average velocity dispersion of 0.3 km $s^{-1}$ , which are similar to the conditions in Taurus, 
is 17 Myrs\cite{Kroupa1995}.
Therefore, our high resolution ALMA observations of IRAS 04191+1523 imply that 
wide binaries in a large range of mass even down to substellar regime can be formed by turbulent fragmentation
 although parameter spaces for very low mass turbulent cores have not been yet explored by theoretical calculations\cite{Goodwin2004, Fisher2004, Offner2010}.
\clearpage

\noindent{\bf METHODS}

\noindent {\bf ALMA observations} 

IRAS 04191+1523 was observed using the Atacama Large Millimeter/submillimeter Array (ALMA) 
during  Cycle 2  (2013.1.00537.S, PI: Jeong-Eun Lee) on 2015 Jan. 4 
and during Cycle 3 (2015.1.00186.S, PI: Michael Dunahm) on 2016 Sep. 28.
For the Cycle 2 observation, three spectral windows were centered  at 219.5604 GHz (\cooo\ \jj21), 
216.113 GHz (DCO$^+$ \jj32), and 231.322 GHz (N$_2$D$^+$ \jj32), 
each with a bandwidth of 58.6 MHz and spectral resolution of 61 kHz ($ \Delta v \sim$0.08 km s$^{-1}$). 
A continuum window was centered at  234 GHz with a bandwidth of 468.75 MHz.
The phase center was at $(\alpha, \delta)_{J2000}=(04^{h}22^{m}00.41^{s}, +15\degree30'21.2'')$, 
and the total observing time was 54 minutes. 
Thirty-eight 12-m antennas were used with baselines in the range from 15 m ($\simeq$10 k$\lambda$) 
to 350 m ($\simeq$255 k$\lambda$) to provide the synthesized beam size of $1''.83 \times 1''.26$ (PA=$-88.3^{\circ}$)
when the natural weighting is adopted.

For the Cycle 3 observation, two continuum windows were centered at 231.1 GHz and 218. 5 GHz with a bandwidth of 2 GHz.
Two spectral windows were set to cover CO \jj21, $^{13}$CO \jj21, and C$^{18}$O \jj21, 
but the total observing time (13 minutes) was too short to detect those lines.
The phase center was at $(\alpha, \delta)_{J2000}=(04^{h}22^{m}00.079^{s}, +15\degree30'24.833'')$.
Forty 12-m antennas were used with baselines in the range from 15 m ($\simeq$11 k$\lambda$) to 3 km ($\simeq$2327 k$\lambda$)
to provide the synthesized beam size of $0''.24 \times 0''.14$ (PA=$14^{\circ}.7$) when the uniform weighting is adopted.

We carried out a standard reduction using CASA\cite{McMullin2007} for both data sets.
The nearby quasar J0510+1800 was used for phase and flux calibration, and J0423-0120 provided a bandpass calibration 
for the Cycle 2 data.
For the Cycle 3 data, J0510+1800 was used for bandpass and flux calibration while J0407+0742 was used for phase calibration. 
The visibility data were imaged using the CLEAN algorithm with natural weighting for \cooo\ to have a higher signal-to-noise ratio 
while uniform weighting was used for the continuum images to provide higher resolutions.
For the Cycle 2 data, the RMS noise levels for C$^{18}$O, DCO$^+$, and N$_2$D$^+$ are 7, 5, and 8 mJy beam$^{-1}$, respectively, 
while the continuum rms noise is 1.0 mJy beam$^{-1}$.
For the Cycle 3 data, the rms noise level for the continuum is 0.45 mJy beam$^{-1}$.

\noindent {\bf Disk masses and inclinations} 

In the 1.3 mm continuum, the disks are unresolved at the lower resolution of $\sim$1.2\arcsec\ while they are resolved 
at the high resolution of $\sim$0.2\arcsec\ (see Figure 2 in the main paper). 
The disk of A is marginally resolved and its deconvolved size is about 26 AU
in diameter. The deconvolved disk size of B is about 85 AU in diameter, which is much bigger than the beam size.
We use the high resolution images to derive disk properties.
Supplementary Table 1 shows the results of Gaussian fitting of the continuum emission.
When the emission is optically thin and the dust temperature is constant, the disk mass is calculated by
\begin{equation}
M_{\rm dust} = \frac{D^2 F}{\kappa_{1.3} B(T_{\rm dust})},
\end{equation}
where $D$ (=140 pc) is the distance to the source, $F$ is the flux density at 1.3 mm, and B is the Planck function 
at the dust temperature, $T_{\rm dust}$ at 1.3 mm.
The assumed dust opacity, commonly used for disks\cite{Andrews2005},
is $\kappa_{1.3}$= 3.5~cm$^2$~g$^{-1}$ .

The deconvolved disk sizes and the inclination of each disk
are listed in Supplementary Table 2. 
The inclination of a disk can be calculated by the size ratio between major and minor axes, 
$cos^{-1}\left({\theta}_{\rm min}/{\theta}_{\rm maj}\right)$.
The calculated inclinations for A and B are $\sim55\degree$ and $\sim59\degree$, respectively.

\noindent {\bf The C$^{18}$O  emission}

Supplementary Figure 1 shows the C$^{18}$O spectra extracted  from 
the two disk positions (A and B) and a position (C) in between as marked in Figure 1 of the main paper.
The broad wings detected in C$^{18}$O  trace the disk rotation.
The red and blue wing components are shaded as red and blue, where the emission is 
integrated to demonstrate the disk rotation in Figure 3 of the main paper. 
For IRAS 04191+1523 B, the blue wing spectrum is flipped on top of the 
red wing spectrum to show that the red wing is contaminated by an additional component,
 which is likely the emission from infalling material because \cooo\ \jj21 is a high density tracer.
Therefore, we use only the blue wing to determine the Keplerian rotation for protostar B.

The velocity profiles for high velocity wings of IRAS 04191+1523 A and B, respectively, along the white lines in Figure 3 
of the main paper are presented in Supplementary Figure 2 in logarithmic scale. 
In this figure, the X-axis represents the position of peak emission found by Gaussian fitting of the emission distribution 
along the white line in Figure 3 at a given velocity. 
The best-fit power-law profiles to the data have power indices similar to the value (0.5) for the Keplerian rotation;
the scatter could be caused by/from contamination of other gas motions such as infall. 
As a result, we applied the Keplerian rotation to their velocity profiles to derive the masses of two protostars 
(see Figure 4 in the main paper.) The masses after correction for disk inclination are 0.14 and 0.12 \msun, 
for protostars A and B, respectively (see Supplementary Table 3). With these masses, we can test the stability of this protostellar binary system\cite{Pineda2015}.
When the velocity difference of two sources is $\sim$0.1 km s$^{-1}$ 
(similar to the velocity resolution of the C$^{18}$O observation) and the projected separation is considered as the actual separation,
the ratio of kinetic energy to gravitation energy is about 0.04 and 0.05 for protostars A and B, respectively, indicating
that this binary is likely bound.

The projected position angles of two disks were determined using channel maps as presented in Supplementary Figure 3.
The blue- and red-shifted components from the source velocity ($v_s$) by the  amounts indicated above each panel are presented as contours.
In each channel, the positions of the blue- and red-shifted emission peaks were connected to determine the position angle.
The mean and standard deviation of the position angles for velocity
shifts between $v_s\pm0.9$ and $v_s\pm1.4$ 
are listed in Supplementary Table 1. 
The difference in the projected position angles of the two disks is about 77$\degree$. 
The true position angle difference depends on
projection; if we consider the disk inclinations derived from the continuum images, the position angle difference 
of the two disks in space is $63\pm5.8\degree$ or $98\pm5.5\degree$ depending on the direction of two disk inclinations.

\noindent {\bf Data Availability}

The data that support the plots within this paper and other findings of
this study are available from ALMA archives with project codes
2013.1.00537.S (https://almascience.nao.ac.jp//aq/?project\_code=2013.1.00537.S)
and 2015.1.00186.S (http://almascience.nao.ac.jp//aq/?project\_code=2015.1.00186.S) and from the corresponding author upon reasonable request.

\newcommand{\nat}{{ Nature }}
\newcommand{\aap}{{Astron. \& Astrophys. }}
\newcommand{\aj}{{ Astron.~J. }}
\newcommand{\apj}{{ Astrophys.~J. }}
\newcommand{\araa}{{Ann. Rev. Astron. Astrophys. }}
\newcommand{\apjl}{{Astrophys.~J.~Letters }}
\newcommand{\apjs}{{Astrophys.~J.~Suppl. }}
\newcommand{\apss}{{Astrophys.~Space~Sci. }}
\newcommand{\icarus}{{Icarus }}
\newcommand{\mnras}{{MNRAS }}
\newcommand{\pasp}{{ Pub. Astron. Soc. Pacific }}
\newcommand{\ssr}{{Space Sci. Rev.}}
\newcommand{\planss}{{Plan. Space Sci. }}
\newcommand{\physrep}{{ Phys. Rep.}}
\newcommand{\bain}{{Bull.~Astron.~Inst.~Netherlands }}

\begin{addendum}
 \item  
This paper makes use of the following ALMA data: ADS/JAO.ALMA\#2013.1.00537.S, and 2015.1.00186.S.
ALMA is a partnership of ESO (representing its member states), NSF (USA) and
NINS (Japan), together with NRC (Canada), and NSC and ASIAA (Taiwan), 
and KASI (Republic of Korea), in cooperation with the Republic of Chile. 
The Joint ALMA Observatory is
operated by ESO, AUI/NRAO and NAOJ. 
 J.-E. Lee was supported by the Basic Science Research
Program through the National Research Foundation of Korea (NRF)
(grant No. NRF-2015R1A2A2A01004769)
and the Korea Astronomy and Space Science Institute under the R\&D
program (Project No. 2015-1-320-18) supervised by the Ministry of Science,
ICT and Future Planning. NJE thanks KASI for support
for a visit to participate in this work.

 \item[Author Contributions]  JEL and SL performed the detailed
calculations used in the analysis.  SL and KT reduced the ALMA data.    
JEL wrote the manuscript.  All authors
were participants in the discussion of results, determination of the
conclusions, and revision of the manuscript.

 \item[Competing Interests] The authors declare that they have no
competing financial interests.
 \item[Correspondence] Correspondence and requests for materials
should be addressed to Jeong-Eun Lee~(email: jeongeun.lee@khu.ac.kr).

\end{addendum}


\clearpage


\renewcommand{\figurename}{{\bf Supplementary Figure }}
\begin{figure*}
\label{fig:s1}
\includegraphics[height=10cm]{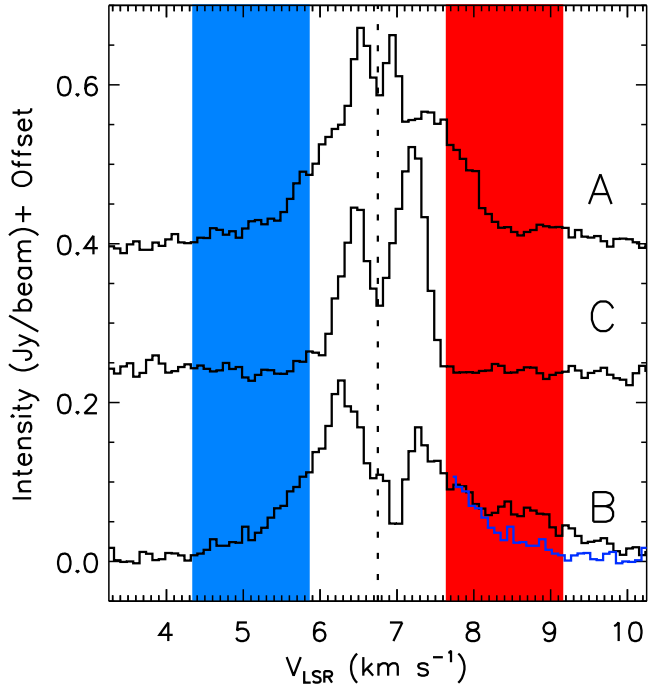}
\caption{ The C$^{18}$O spectra for the two disk positions and in between the two disks (position C in figure 1 of the main paper). The spectrum at  C represents the common envelope with no disk emission. 
Therefore, we define the velocity ranges (shaded as blue and red) for the disk emission based on this spectrum.  
The blue line above 7.65 km s$^{-1}$ at the position B indicates the  spectrum of the blue wing reflected about the source velocity of 6.75 km s$^{-1}$, which is marked with the dashed line.
It shows an additional bump in the red wing at the position B. 
A Gaussian fit places the additional component at 8.9 km s$^{-1}$.}
\end{figure*}

\begin{figure*}
\label{fig:s2}
\includegraphics[height=5cm]{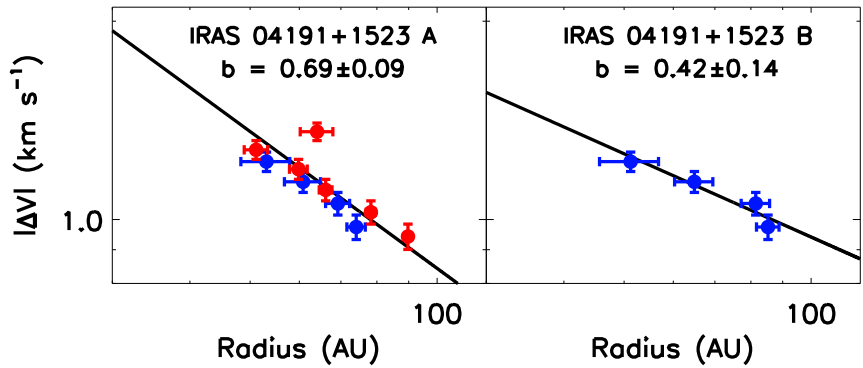}
\caption{ The velocity profiles for the high velocity wings of IRAS 04191+1523 A and B, respectively, 
in logarithmic scale. The data points and error bars are the same as in Figure 4. 
When the velocity profile follows a power-law function with radius
 ( $|\Delta V| \propto r^{-b}$), the least squares fitting method show the best-fit power indices (and their 
 uncertainties) of A and B are 0.69 ($\pm$0.09) and
0.42 ($\pm$0.14), respectively. These values are similar to the power index (0.5) for the Keplerian rotation.}
\end{figure*}

\begin{figure*}
\label{fig:s3}
\includegraphics[height=10cm]{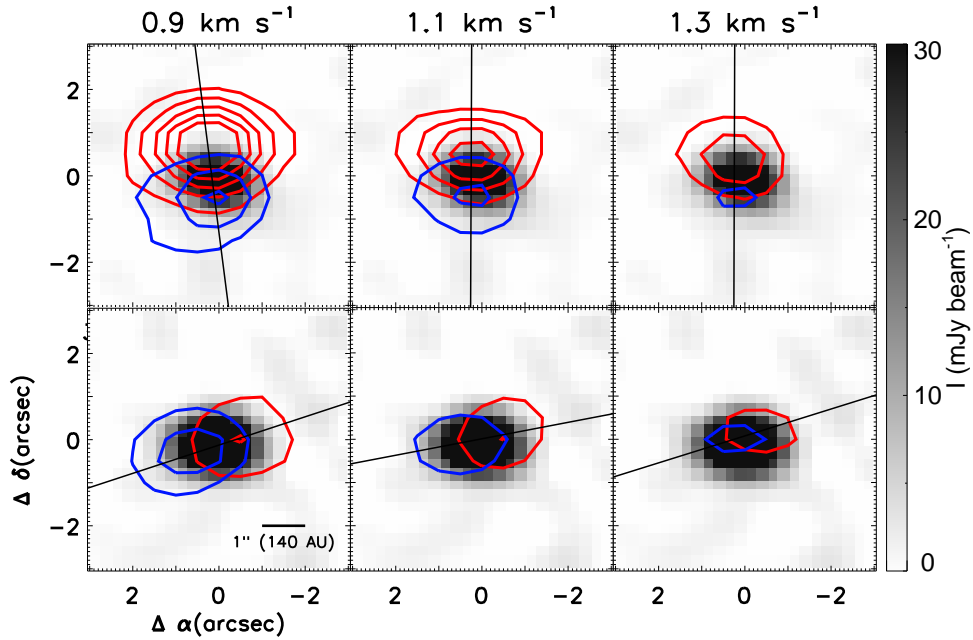}
\caption{ The channel maps of high velocity components of the C$^{18}$O emission 
for IRAS 04191+1523 A (top) and B (bottom) on top of the continuum image (gray scale).
The lowest contour and subsequent contour step are 5 times the rms noise, which is 6.9 mJy beam$^{-1}$ 
and 10.9 mJy beam$^{-1}$ for A and B, respectively.
The thin black line marked in each box connects the blue- and red-shifted emission peaks. 
The mean and standard deviation of the position angles measured from the thin black lines are $2.9\pm4.1\degree$ 
and $-74.0\pm6.1\degree$.
The velocity marked at the top of each column indicates the velocity shifted from the source
velocity. }
\end{figure*}

\clearpage

\begin{center}
\begin{tabular}{cccccc}
\multicolumn{6}{c}{\bf Supplementary Table 1: Source Properties of IRAS 04191+1523}
\label{tb:gauss_cont}
\\\hline
{} &
{\bf R.A.} &
{\bf Dec.} &
{Continuum Flux density} &
{Continuum Peak Intensity} &
{Disk mass$^a$} \\ 
{} &
{} &
{} &
{mJy} &
{mJy beam$^{-1}$} &
{$\times 10^{-3}$ \msun}
\\\hline\hline
 A &  04:22:00.4283  & +15:30:21.189 &  53.5 & 33.1  & 3.70  \\
   & (0$''$.001)     & (0$''$.002)   & (1.1)& (0.46)& (0.08) \\
B  &  04:22:00.0826  & +15:30:24.605 &  71.1 & 10.2 &  4.91 \\
   & (0$''$.001)     & (0$''$.009)   & (3.2)& (0.41)& (0.22)\\
\hline
\multicolumn{6}{l}{\footnotesize The continuum flux density and peak intensity are measured with the task, imfit in CASA. 
The value in parenthesis is the error (1 $\sigma$).} \\
\multicolumn{6}{l}{\footnotesize $^a$The disk mass is calculated by the equation (1) in Methods.}\\
\end{tabular}
\end{center}

\clearpage
\begin{tabular}{ccccc}
\multicolumn{5}{c}{\bf Supplementary Table 2: Deconvolved Disk Sizes and Inclinations of IRAS 04191+1523}
\label{tb:disk_inc}
\\\hline
{} &
{${\theta}_{\rm maj}$ $^a$} &
{${\theta}_{\rm min}$ $^a$} &
{${\theta}_{\rm incl.}$$^b$} &
{${P.A.}$} \\

{} &
{mas} &
{mas} &
{$^\circ$} &
{$^\circ$}
\\\hline\hline
A &  189.3 (9.0)  & 109.4 ( 5.4) & 54.6 (2.8) & 0.0 (3.9)  \\
B &  608 (29)     & 309 (20)     & 59.3 (2.7) & -55.2 (3.1)  \\
\hline
\multicolumn{5}{l}{\footnotesize These are measured with the CASA task, imfit. 
The value in the parenthesis is the error (1 $\sigma$). } \\
\multicolumn{5}{l}{\footnotesize  The synthesized beam is 0$''$.24$\times$0$''$.14 (P.A. = 14.72). } \\
\multicolumn{5}{l}{\footnotesize $^a$FWHM of major and minor axes. }\\
\multicolumn{5}{l}{\footnotesize $^b$  ${\theta}_{\rm incl.}$ = $cos^{-1}\left({\theta}_{\rm min}/{\theta}_{\rm maj}\right)$.}
\end{tabular}

\clearpage

\begin{tabular}{ccc}
\multicolumn{3}{c}{\bf Supplementary Table 3: Protostellar mass derived by Keplerian rotation velocity profile}
\label{tb:stellar_mass}
\\\hline
{} &
{Minimum mass$^a$} &
{Actual mass$^b$} \\
{} &
{\msun} &
{\msun} 
\\\hline\hline
A &  0.09 (0.003)    & 0.14 (0.01) \\
B &  0.09 (0.006)    & 0.12 (0.01) \\
\hline
\multicolumn{3}{l}{\footnotesize $^a$ Protostellar mass derived assuming edge-on disk.}\\
\multicolumn{3}{l}{\footnotesize $^b$ Protostellar mass corrected for inclination (see Supplementary Table 2).}
\end{tabular}

\end{document}